**Polymer-Shell Coating of Mie-Resonant Silicon Nanospheres for Controlled Fabrication of Self-Assembled Monolayer**

Oanh Vu, Jialu Song, Hiroshi Sugimoto*, and Minoru Fujii*

Department of Electrical and Electronic Engineering, Graduate School of Engineering, Kobe University, Rokkodai, Nada, Kobe 657-8501, Japan

E-mail: sugimoto@eedept.kobe-u.ac.jp

E-mail: fujii@eedept.kobe-u.ac.jp

**Abstract.**

A polymer shell offers a unique opportunity to tailor structural and optical properties of optically functional nanoparticles and their ensembles. Here, we develop a process to coat Mie-resonant silicon nanospheres (Si NSs) with a thermoresponsive poly(N-isopropylacrylamide) (PNIPAM) hydrogel shell. We show formation of a PNIPAM shell of controlled thickness from the change of the hydrodynamic diameter and study the effect of thermoresponsive shrinkage and expansion of the shell on the Mie resonance of a Si NS. We then demonstrate that Si NSs with PNIPAM shells enable fabrication of cluster-free, self-assembled monolayers of Si NSs, in which distances between NSs are controlled by the PNIPAM shell thickness.

# 1. Introduction

In recent years, dielectric nanoantennas have attracted increasing attention in the field of nanophotonics and metamaterials due to their ability to enhance light-matter interactions through the Mie resonances.[1–5] Unlike their metallic counterparts, dielectric nanoantennas do not suffer from ohmic losses, allowing for more efficient operation in nonlinear optics, biosensing, and bioimaging.[6–9] Silicon (Si) is one of the most widely employed materials in dielectric nanophotonics because of the high refractive index ($n\sim4$) and the low optical loss.[5,10–13] In addition to the superior material property, the high compatibility with modern nanofabrication processes bestows a special position to Si in a nanophotonic society. However, nanofabrication processes, especially electron beam lithography, are still expensive for large area devices, and thus more cost-effective production processes of Si-based nanoantennas and metamaterials are highly desired.

In the case of plasmonic materials, the bottom-up assembly has often been employed for the fabrication of nanophotonic devices by using colloidal suspensions of gold and silver nanoparticles as precursors.[14–20] In principle, the same processes can be applied to dielectric nanoparticles if the colloidal suspensions are available. Our group has developed colloidal suspensions of spherical Si nanoparticles (Si nanosphere (NS)) having the Mie resonances in the visible to near infrared range.[3,10,21] The suspensions exhibit vivid and size-dependent scattering colors arising from Mie resonances of individual Si NSs.[10,21] From the NS suspensions, Si NS monolayers, linear arrays, zig-zag arrays, 2-dimensional square lattice arrays, etc. have been produced by a bottom-up assembly process such as a template-assisted self-assembly method and a Langmuir–Blodgett (LB) method.[22–24] These Si NS-based nanoantennas and metasurfaces exhibit collective Mie resonances depending on the NS alignment. For example, Si NS monolayers produced by the LB method exhibit size-dependent reflection colors with the peak reflectance exceeding 30%.[24] A major challenge in a bottom-up assembly process is statistical distribution of the NS-to-NS distance and unavoidable formation of the clusters such as the dimers and trimers even at a relatively low NS density. NS clusters exhibit optical responses significantly different from those of individual ones. They often have broader resonances and degrade the performance of fabricated metasurfaces.[22,25]

In this work, we develop colloidal Si NSs that are compatible with bottom-up assembly processes, but can avoid NS clustering. Our strategy is the formation of a polymer shell on a Si NS core. In the core–shell architecture, the minimum NS-to-NS distance is defined by the shell thickness, allowing control of the coupling strength between NSs.[26–29] We first show that a uniform and thickness-controlled hydrogel shell of poly(N-isopropylacrylamide) (PNIPAM) can be formed on the surface of a Si NS core (Si@PNIPAM particle) in a wet-chemical process. We then fabricate monolayers of Si@PNIPAM particles with different thickness PNIPAM shells on a silica substrate using a LB method. We demonstrate that cluster-free Si NS monolayers with controlled NS densities can be produced.

## 2. Results and Discussion

### 2.1 Fabrication and characterization of Si@PNIPAM particle

To fabricate Si@PNIPAM particles, we first prepare colloidal suspensions of Si NSs *via* thermal disproportionation of silicon monoxide (SiO).[10,21] Details of the fabrication process are shown in the Experimental Section. **Figure 1a** shows a photo of methanol solution of Si NSs. The solution exhibits a brownish color due to the presence of different size Si NSs. **Figure 1b,c** shows TEM images of Si NSs. The Si NSs are nearly perfectly spherical and have high crystallinity. **Figure 1d** (solid curve) shows the measured backward scattering spectrum of a single Si NS with the diameter of 195 nm on a $SiO_2$ substrate (see **Methods**), together with the calculated spectrum (dotted curve) that takes into account the substrate and the numerical aperture of the collection optics. The measured spectrum agrees very well with the simulated one. This guarantees the high quality of the Si NS. The peaks are assigned to the magnetic dipole (MD), electric dipole (ED), magnetic quadrupole (MQ), and electric quadrupole (EQ) resonances.

PNIPAM shells are formed on these Si NSs following the procedure shown in **Figure 1e**.[30] Si NSs are first functionalized with 3-Methacryloxypropyltrimethoxysilane (MPS) to enable covalent anchoring of the polymer shell, and then PNIPAM shells are grown onto the MPS-functionalized Si NSs *via* radical polymerization in a batch process. **Table 1** summarizes the shell formation parameters. In all conditions ((1)-(7)), the amount of Si NSs is fixed at 120 mg/100 mL. In condition (1)-(5), the concentration of the monomer N-isopropylacrylamide (NIPAM) is changed from 10 mM to 50 mM, while the ratios of the crosslinker *N,N'*-methylenebis(acrylamide) (BIS) and the free-radical initiator ammonium persulfate (APS) are fixed to 10 mol% vs. NIPAM and 4 mol% vs. NIPAM, respectively. In conditions (6) and (7), the concentration of NIPAM and the ratio of the APS free-radical initiator are fixed, while the ratio of BIS crosslinker is changed.

**Table 1.** PNIPAM shell formation condition. The amount of Si NSs is fixed at 120 mg/100 mL for all conditions.

| PNIPAM shell formation condition | NIPAM (mM) | BIS (mol% vs. NIPAM) | APS (mol% vs. NIPAM) |
|---|---|---|---|
| (1) | 10 | 10 | 4 |
| (2) | 15 | 10 | 4 |
| (3) | 25 | 10 | 4 |
| (4) | 40 | 10 | 4 |
| (5) | 50 | 10 | 4 |
| (6) | 25 | 2 | 4 |
| (7) | 25 | 5 | 4 |

In an as-prepared Si@PNIPAM particle suspension, different diameter particles are contained. Si@PNIPAM particles are size-separated by a density-gradient centrifugation process (**Figure 1f**).[10,21,31] **Figure 1g** shows photographs of 14 size-separated fractions obtained from an original Si@PNIPAM suspension, and **Figure 1h** shows the extinction spectra of the suspensions. We estimate the average diameter of Si NS core ($D_{core}$) and the standard deviation ($\sigma$) by fitting the extinction spectra with calculated extinction cross-section spectra assuming a normal size distribution (**Supporting Information S1**).[10,21] The estimated $D_{core}$ and $\sigma$ are indicated below the photographs. It is important to note that a PNIPAM shell does not affect the scattering spectrum of a Si NS in water because its refractive index is very close to that of water, i.e., extinction spectra of a bare Si NS and a Si@PNIPAM particle in water are very similar, and thus the size estimated from the extinction spectrum corresponds to the diameter of the Si NS core in a Si@PNIPAM particle.

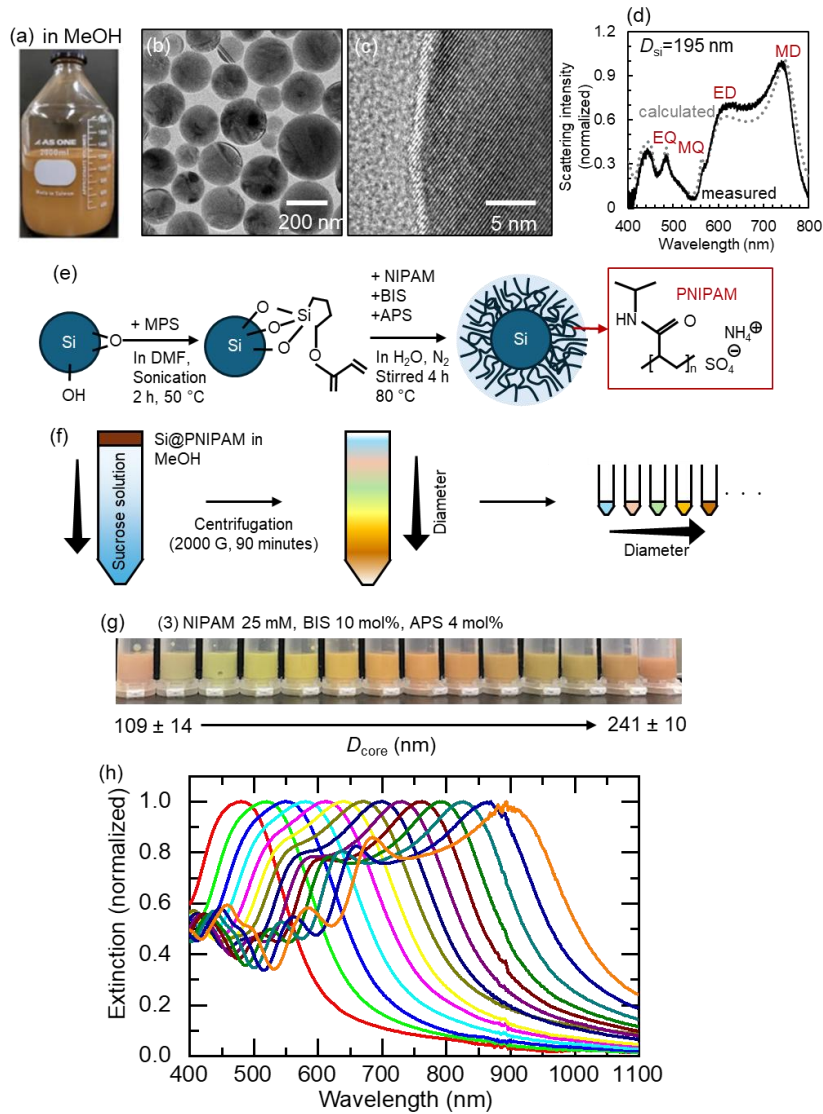

**Figure 1.** (a) Photograph of a 1L bottle of a methanol suspension of Si NSs. (b, c) TEM images of Si NSs. Lattice fringe in (c) corresponds to the {111} plane of crystalline Si. (d) Measured (solid curve) and calculated scattering spectra (dotted curve) of a single Si NS with 195 nm in diameter on $SiO_2$ substrate. Peaks of the MD, ED, MQ and EQ Mie resonances are designated. (e) Schematic illustration of PNIPAM shell formation process. (f) Schematic illustration of density-gradient centrifugation process for size separation of Si@PNIPAM particle suspension. (g) Photographs of water suspensions of size-separated Si@PNIPAM particles. The diameter of Si NS core ($D_{core}$) is changed from 109 ± 14 nm to 241 ± 10 nm. (h) Extinction spectra of Si@PNIPAM suspensions in (g).

In order to estimate the whole diameter of a Si@PNIPAM particle, we measured the hydrodynamic diameter ($D_h$). In **Figure 2a**, $D_h$ is plotted as a function of $D_{core}$ for Si@PNIPAM particles produced with conditions (1)-(5) in **Table 1**. The data for bare Si NSs are also shown as a reference (brown). In bare Si NSs, $D_h$ is slightly (~5-30 nm) larger than $D_{core}$, especially in the small $D_{core}$ range. This is due to the formation of a hydration layer around a hydrophilic Si NS in water. By the formation of a PNIPAM shell, $D_h$ increases drastically. From the values of $D_h$ and $D_{core}$, the PNIPAM shell thickness ($t_s = \frac{D_h - D_{core}}{2}$) can be estimated. **Figure 2b** shows $t_s$ as a function of NIPAM concentration for particles with $D_{core}$=200 nm. The shell thickness increases from 19 to 210 nm when the NIPAM concentration increases from 10 to 50 mM. **Figure 2c** shows the $D_h$ - $D_{core}$ relation for Si@PNIPAM particles produced with conditions (3), (6) and (7) in **Table 1**, and **Figure 2d** shows $t_s$ obtained from **Figure 2c** for particles with $D_{core}$=200 nm. $t_s$ increases with increasing the ratio of crosslinker BIS.

**Figure 2e,f** shows TEM images of dried Si@PNIPAM particles produced with condition (2) and (5), respectively. We can see a blurry shell of PNIPAM around a Si NS core. The shell thickness is 18 nm and 166 nm, respectively. These values are slightly smaller than those estimated from **Figure 2c**. This is because PNIPAM is in the swollen state in water.[32]

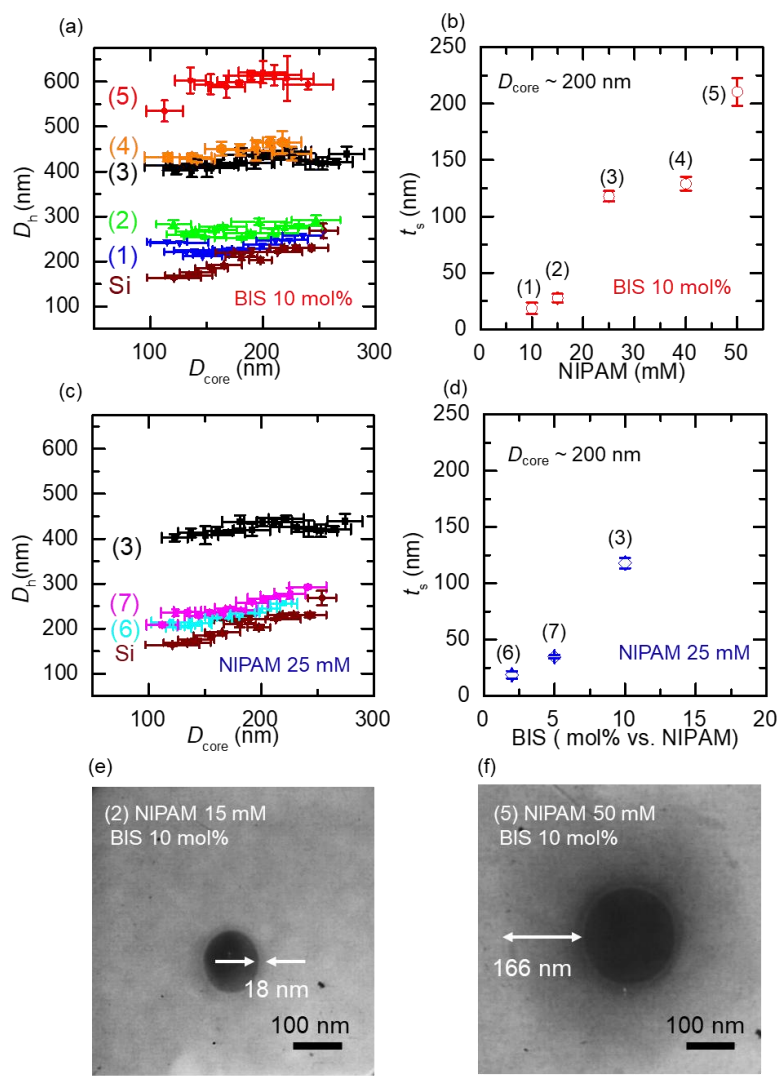

**Figure 2.** (a) Hydrodynamic diameter ($D_h$) of Si@PNIPAM particles measured at 25 °C as a function of Si NS core diameter ($D_{core}$). The data of Si@PNIPAM particles produced with conditions (1)-(5) in **Table 1** are shown. The data of bare Si NSs are also shown as a reference (brown). (b) PNIPAM shell thickness ($t_s = \frac{D_h - D_{core}}{2}$) obtained from (a) as a function of NIPAM concentration for particles with $D_{core}$=200 nm. (c) $D_h$ as a function of $D_{core}$ for Si@PNIPAM particles produced with conditions (3), (6) and (7). (d) $t_s$ obtained from (c) as a function of the ratio of BIS crosslinker for particles with $D_{core}$=200 nm. (e, f) TEM images of dried Si@PNIPAM particles produced with condition (2) (e) and (5) (f).

## 2.2 Thermoresponse of Si@PNIPAM particles

PNIPAM is a thermoresponsive polymer, which exhibits a temperature-dependent volume phase transition (VPT) around its lower critical solution temperature (LCST) approximately 32 °C.[33] Below the LCST, PNIPAM absorbs a large amount of water and is at the swollen state (water-rich phase), while above the LCST, it repels water leading to shrinkage and becomes the polymer-rich phase (**Figure 3a**).[34] In Si@PNIPAM particles, this transition can be monitored from the $D_h$. **Figure 3b** shows $D_h$ as a function of temperature for Si@PNIPAM particles prepared with conditions (1)-(5). $D_{core}$ is shown in the figure. $t_s$ obtained from **Figure 3b** is plotted in **Figure 3c**. In **Figure 3b** and **c**, we can see that the PNIPAM shell collapses into a compact state around the LCST. For example, in condition (5), $t_s$ changes from 200.4 nm to 79.5 nm when the temperature increases from 20 °C to 50 °C.

The shrinkage of the shell modifies the surrounding dielectric environment of a Si NS and may affect the Mie resonance property. **Figure 3d** shows the extinction spectra of a water suspension of Si@PNIPAM particles ($D_{core}$=210 nm, σ=29.3 nm, condition (5)). The temperature is changed from 20 to 50 °C. The spectrum, especially the long wavelength range around the MD resonance, changes with temperature. In the inset, the peak wavelength of the MD resonance is plotted as a function of the temperature. It jumps at the LCST. The observed temperature dependence of the spectrum is reversible **(Supporting Information S2)**.

To explain the observed effect, we calculate the extinction cross-section spectra for a Si@PNIPAM particle with $D_{core}$=210 nm for two cases. In the water-rich phase, we assume that the refractive index of the shell is the same as that of water and thus the shell does not exist for light. In the polymer-rich phase, the refractive index is assumed to be 1.5 and the experimentally obtained value at 50 °C is employed for $t_s$ (79.5 nm). The calculated extinction spectra are shown in **Figure 3e**. The experimentally observed red-shift of the MD resonance at high temperature is reproduced by the model, although it is too crude to reproduce the whole spectral shape. Similar data obtained for a Si@PNIPAM particle with $D_{core}$=123 nm is shown in **Supporting Information S3**.

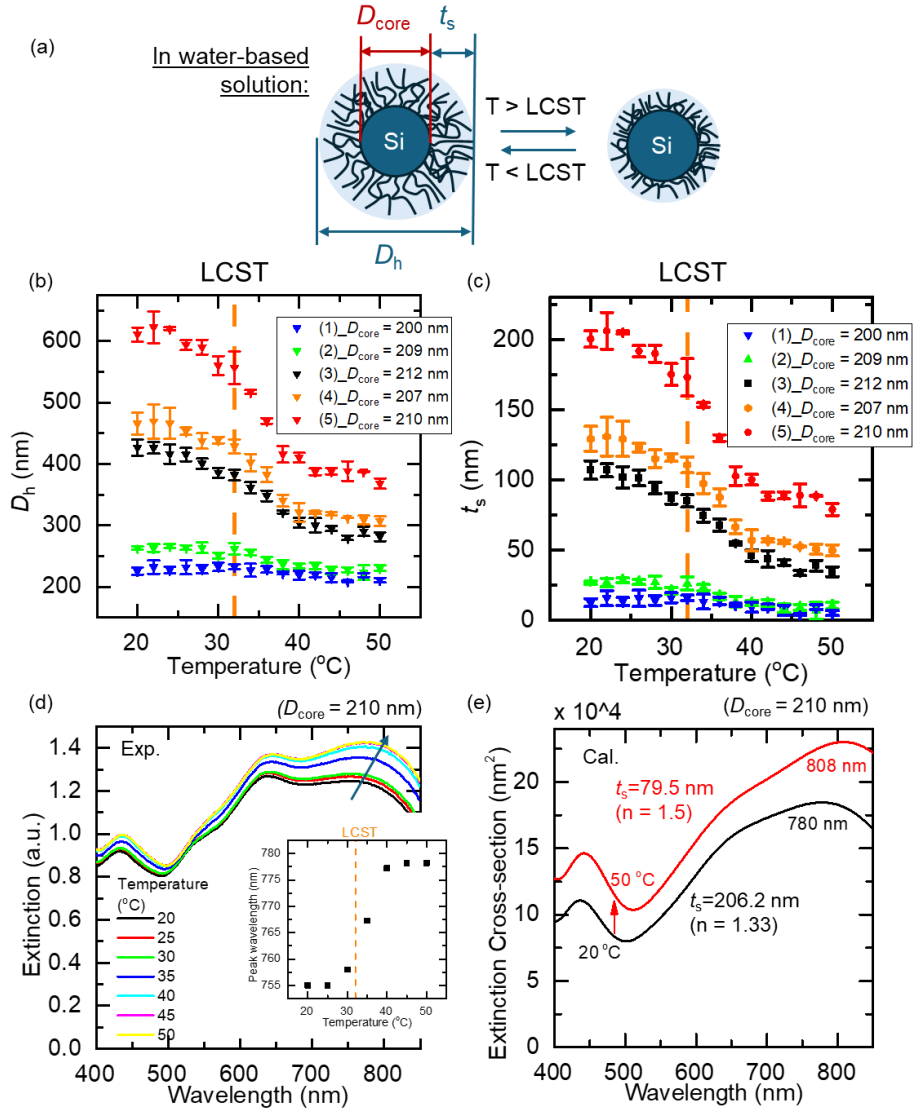

**Figure 3.** (a) Schematic illustration of a Si@PNIPAM particle in a water-based solution below and above the LCST. (b) Hydrodynamic diameter ($D_h$) of Si@PNIPAM particles as a function of temperature. Conditions for PNIPAM shell formation are (1) to (5). $D_{core}$ is shown in the figure. (c) Shell thickness ($t_s$) obtained from (b) as a function of temperature. (d) Extinction spectra of Si@PNIPAM particle suspension at different temperatures (20–50 °C). The condition for PNIPAM shell formation is (5) and $D_{core}$ is around 210 nm. In the inset, MD resonance peak wavelength is plotted as a function of temperature. (e) Calculated extinction cross-section spectra of a Si@PNIPAM particle below and above the LCST. $D_{core}$ is 210 nm. The water-rich phase below the LCST is modeled with the shell of the refractive index same as that of water ($n$=1.33) (black curve),

while the polymer-rich phase above LCST is modeled with a 79.5 nm-thick shell of $n$=1.5 (red curve). The shell thickness is estimated from experimental data.

## 2.3 Formation of cluster-free Si NS monolayer

PNIPAM shells in Si@PNIPAM particles serve as steric barriers between particles, enabling formation of cluster-free self-assembled monolayers of Si NSs. **Figure 4a** shows the procedure to produce a Si NS monolayer from Si@PNIPAM particle suspensions by the LB method. Colloidal dispersion of Si@PNIPAM particles is gently transferred to the air/water interface. Spontaneously self-assembled particles are then transferred onto a SiO$_2$ substrate and dried. **Figure 4b** shows the cross-sectional scanning electron microscope (SEM) image of a Si@PNIPAM monolayer. The dome-like shape indicates that Si NSs are covered by dried PNIPAM shells on the substrate. The shells can be removed by an ultraviolet (UV)-ozone treatment. **Figure 4c** shows the SEM image after removing the shells. PNIPAM shells are totally removed. It should be noted that the dried PNIPAM shell affects the optical response of a Si NS and thus removing it is essential to obtain distinct Mie scattering identical to that of a bare Si NS. This can be seen by comparing scattering spectra before and after removing a PNIPAM shell. **Figure 4d** compares scattering spectra of a single Si NS with a PNIPAM shell ($D_{core}$=165 nm) before (black) and after (red) PNIPAM shell removal. As a reference, the spectrum of a pristine Si NS ($D_{core}$=165 nm) is also shown (blue). The spectrum after PNIPAM removal is identical to that of the pristine Si NS.

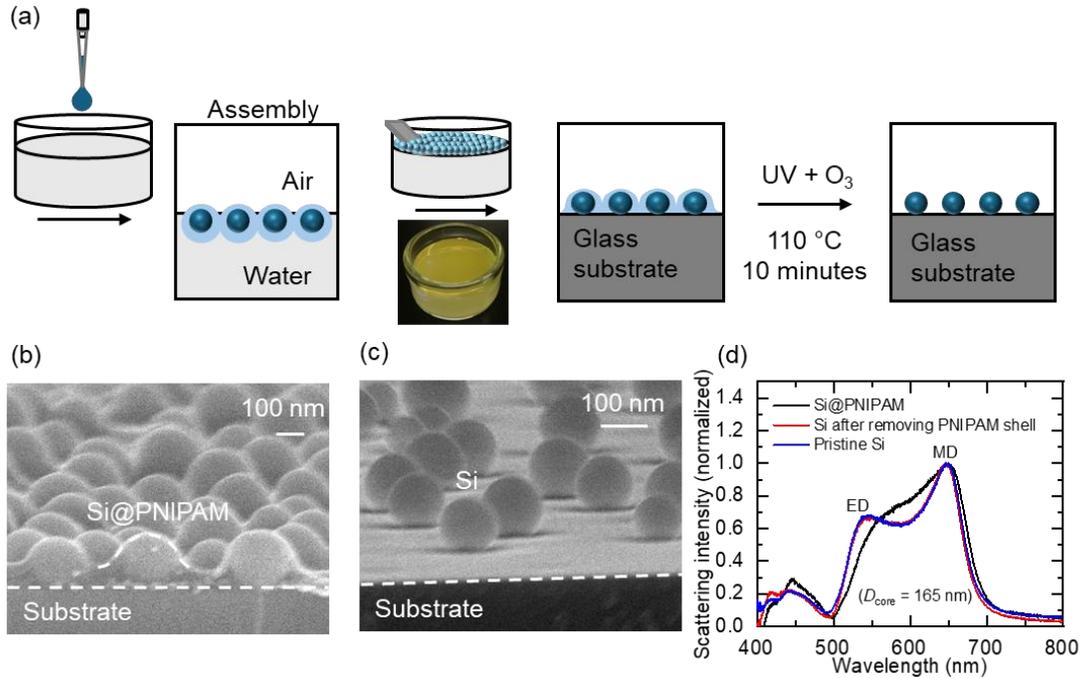

**Figure 4.** (a) Schematic illustration of the procedure to fabricate Si NS monolayer on SiO$_2$ substrate. (b) Cross-sectional SEM image of Si@PNIPAM particles on SiO$_2$ substrate. $D_{core}$ is around 130 nm and PNIPAM shell formation condition is (2). The process temperature is below the LCST. (c) Cross-sectional SEM image of Si@PNIPAM particles on SiO$_2$ substrate after UV-O$_3$ treatment. PNIPAM is removed and bare Si NSs are separately placed on the substrate. (d) Scattering spectra of a single Si@PNIPAM particle before (black curve) and after PNIPAM removal (red curve). $D_{core}$ is 165 nm. As a reference, the spectrum of a pristine Si NS with the same size is shown (blue curve).

**Figure 5a** shows the photographs of Si NS monolayers on SiO$_2$ substrates produced by the process. $D_{core}$ is changed from 112 to 210 nm and the condition for PNIPAM shell formation is fixed to (5). We can see vivid structural colors from blue to orange. Corresponding total reflectance spectra (*i. e.*, the sum of the specular and diffuse reflectance spectra) are shown in **Figure 5b**. We can see MD and ED (MQ) resonances clearly. The spectra are similar to the extinction spectra in **Figure 1h**. This suggests that clustering of Si NSs during the monolayer formation process is effectively prevented by the PNIPAM shell, and Mie resonance properties of individual Si NSs are well preserved on substrates. It should be noted that in Si NS monolayers made from bare Si NSs, these resonance features are broadened due to the inter-NS coupling, and the MD and ED resonances are no longer distinguishable.[24]

The largest advantage of the developed process for the formation of a Si NS monolayer is the capability to control the minimum NS-to-NS distance by the PNIPAM shell thickness. **Figure 5c-f** and **Figure 5g-j**, respectively, show optical microscope and SEM images of Si NS monolayers prepared with the PNIPAM shell formation conditions (5) to (2) and almost fixed $D_{core}$ (~150 nm). The exact values of $D_{core}$ and $D_h$ are shown in the figures. For the formation of these Si NS monolayers by the LB method, almost the same number of Si@PNIPAM particles are dropped on water. Therefore, the particle density is not intentionally controlled by the amount of particles. In **Figure 5c**, we can see densely-packed partially hexagonally-aligned green and yellow dots. In the corresponding SEM image in **Figure 5g**, Si NSs are almost uniformly dispersed on the substrate with large NS-to-NS distances. Therefore, the green and yellow dots in **Figure 5c** arise from Mie scattering of individual Si NSs. In fact, the total reflectance spectrum in **Figure 5l** (yellow) is similar to the backward scattering spectrum of a single Si NS. Mixing of the green and yellow dots in **Figure 5c** is due to the residual size distribution. In the fast-Fourier transform (FFT) image in the inset of **Figure 5g**, a hexagonal pattern can be seen. Therefore, a monolayer composed of well-separated and hexagonally-aligned Si NSs are produced by the process. The interparticle distance estimated from the FFT image is ~520 nm, which is close to $D_h$ (594 nm) in water. The Si NS density estimated from SEM images is 374 /100 μm$^2$ (**Figure 5k**).

The Si NS surface density can be increased by reducing the PNIPAM shell thickness. In **Figure 5d** and **h**, the surface density increases to 660 /100 μm$^2$. However, the shape of the reflectance spectrum in **Figure 5l** (green) does not change and only the intensity is increased. This indicates that the NS-to-NS interaction is still negligibly small, and the reflectance spectrum is the simple sum of the backward scattering spectrum of individual Si NSs. The situation changes by further reducing the PNIPAM shell thickness (**Figure 5e** and **i**). Although the SEM image (**Figure 5i**) does not change so much from that in **Figure 5h**, the reflectance spectrum starts to be modified. The MD and ED resonances become broad due probably to increased NS-to-NS interactions.

When the PNIPAM shell thickness is reduced to ~50 nm, the morphology of the layer changes (**Figure 5f** and **j**). In the SEM image in **Figure 5j**, Si NSs are not uniformly distributed, but form 2-dimensional islands. This is due to soft interaction of PNIPAM shells during the assembly process, which bring the NSs closer together.[35] The optical microscope image in **Figure 5f** becomes more yellowish and the reflectance spectrum becomes broad and red-shifts (**Figure 5l** (purple)). These changes suggest enhanced coupling between Si NSs. It should be emphasized that stronger coupling does not mean physical contact between NSs. TEM observations reveal that even when PNIPAM shell thickness is very thin, NSs are still spatially separated (**Supporting Information S4**).

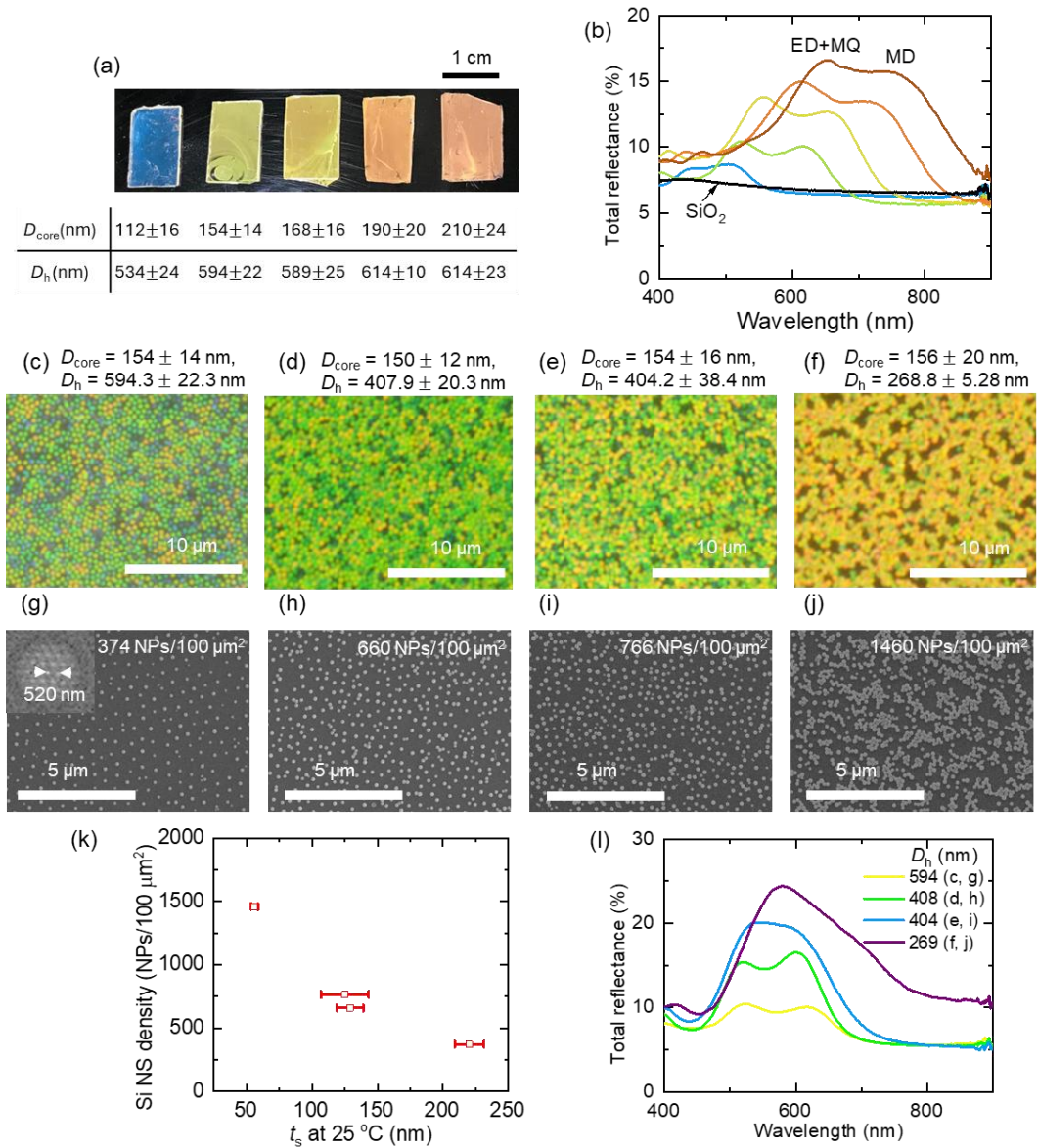

**Figure 5.** (a) Photographs of Si NS monolayers on SiO$_2$ substrates. The PNIPAM shell formation process is condition (5) and $D_{core}$ is changed from 112 to 210 nm. The process temperature is below the LCST. (b) Total reflectance spectra, i.e., the sum of the specular and diffuse reflectance spectra, of Si NS monolayers in (a). As a reference, the total reflectance spectrum of a SiO$_2$ substrate is also shown (black curve). (c-f) Dark-field optical microscope images of Si monolayers. $D_{core}$ is almost fixed (150 nm - 156 nm) and $D_h$ is 594 nm (c), 408 nm (d), 404 nm (e) and 269 nm (f). The corresponding SEM images are shown in (g-j), respectively. The inset in (g) is the FFT image. (k) Number of Si NSs per 100 μm$^2$ on a substrate obtained from SEM images as a function of $t_s$. (l) Total reflectance spectra of Si NS monolayers in (c-f).

## 3. Conclusion

We have developed a process to uniformly coat Mie-resonant Si NSs with PNIPAM hydrogel shells of controlled thickness. We showed that the shell thickness can be modulated by temperature because of the thermoresponsive nature of the PNIPAM shell, and that the Mie resonance property of Si NSs can also be tuned accordingly. The process allowed fabrication of cluster-free self-assembled monolayers of Si NSs from Si@PNIPAM particle suspensions. We showed that the minimum NS-to-NS distance, and consequently the coupling strength, can be controlled by the PNIPAM shell thickness. When the shell is sufficiently thick, the inter-NS coupling is negligibly small, and the reflectance spectrum of a monolayer closely resembles the backward scattering spectrum of individual Si NSs. A reduction in shell thickness enhances interparticle coupling, resulting in the modulation of the reflectance spectrum. We also observed that a PNIPAM shell facilitates alignment of Si NSs on a substrate. A monolayer composed of well-separated and hexagonally-aligned Si NSs is formed when the shell thickness is large. The presented results indicate that Si@PNIPAM particles can be a versatile platform for the production of a variety of Si NS-based photonic architecture with a self-assembled process.

## 4. Experimental Section

**Chemicals.** SiO lump powder (99%, FUJIFILM Wako), hydrofluic acid (HF, 46%, FUJIFILM Wako), hydrogen peroxide (H$_2$O$_2$, 30%, FUJIFILM Wako), 3-(trimethoxysilyl)propyl methacrylate (stabilized with BHT) (MPS, >98%, TCI), *N,N*-dimethylformamide super hehydrated (DMF, FUJIFILM Wako), *N*-isopropylacrylamide (NIPAM, 97%, Sigma), *N,N'*-methylenebis(acrylamide) (BIS, 99%, Sigma), ammonium persulfate (APS, ≥98%, Sigma), methanol (MeOH, 99.8%, FUJIFILM Wako), ethanol (EtOH, 99.5%, FUJIFILM Wako), 2-propanol (IPA, 99.7%, FUJIFILM Wako), acetone (99%, FUJIFILM Wako), deionized water (DI).

**Synthesis of colloidal suspension of Si NS.** Si NSs were prepared by high temperature disproportionation reaction of SiO.[10] SiO lump powder was annealed for 30 minutes at 1500 °C in an $N_2$ gas environment for the disproportionation into Si NSs and $SiO_2$ matrices. Si NSs were then retrieved from $SiO_2$ matrices by etching out the matrices in an HF solution for an hour. After washing with MeOH, Si NSs were dispersed in a MeOH solution.

**Synthesis of Si@PNIPAM particles.** The surface of Si NSs was first functionalized by MPS. Typically, as-produced Si NSs were centrifuged and dried overnight to remove MeOH and then dispersed in DMF superhydrate with the concentration of 0.4 wt%. MPS was added to the solution, then the mixture was sonicated for 2 hours at 50 °C. The amount of added MPS was set to be 500 molecules per Si NS surface of 1 $nm^2$. Finally, Si NSs were washed several times with MeOH by centrifugation.

A PNIPAM microgel shell was polymerized on MPS-functionalized Si NSs via a radical polymerization process. In a 3-neck round-bottom flask, NIPAM and the cross-linker BIS were dissolved in DI. The concentrations of monomer NIPAM and the amount of cross-linker with respect to that of NIPAM are shown in **Table 1**. Then, Si NS dispersed DI was slowly added into the flask under stirring. The mixture was heated to 80 °C using an oil bath stirrer and purged with nitrogen under refluxing. After an equilibration time of 30 minutes, the reaction was started by rapid injection of 1 mL of free-radical initiator APS in water (4 mol% vs. NIPAM monomer) into the flask. The reaction proceeded in 4 hours, then the reaction mixture was allowed to cool to room temperature. The core-shell particles were washed by centrifugation with DI water several times to dispose excess microgels.

**Size separation of Si@PNIPAM particles.** Si@PNIPAM particles were separated by size by a density-gradient centrifugation process.[10,36] A sucrose density gradient solution (20 - 50 wt.%) was first prepared. Then, 700 μL of Si@PNIPAM in MeOH (15 mg/mL) was added onto the top of the sucrose density gradient solution, and centrifuged for 90 minutes at 2000 G (TOMY LCX-100) to produce layers of different size Si@PNIPAM particles. The particles were retrieved layer-by-layer by Piston Gradient Fractionator (BioComp) with the speed of 0.5 mm/s. Size-separated Si@PNIPAM particles were washed several times with DI and stored in DI.

**Formation of Si@PNIPAM monolayer.** A $SiO_2$ substrate (1 × 1.5 $cm^2$ in size) was cleaned by sonication in acetone, IPA and DI water for 10 minutes each, dried by $N_2$ gun, cleaned and hydroxylated by a UV-ozone cleaner (SAMCO, UV-1) for 20 min at 110 ºC. A monolayer of Si@PNIPAM particles was assembled on a $SiO_2$ substrate by the LB method as follows. A small crystallization dish was filled with DI water. 200 μL of Si@PNIPAM particle dispersion (purified, in ethanol) was gently transferred to the air/water interface. The particles spontaneously self-assembled into a freely floating monolayer at the interface. By fully immersing a $SiO_2$ substrate in water under the floating monolayer and then withdrawing it slowly, the monolayer was transferred onto the substrate. The monolayer was dried in air at room temperature. Finally, PNIPAM shells were removed by UV-ozone treatment.

**Characterizations.** Hydrodynamic diameter of Si@PNIPAM particles in DI water was measured by dynamic light scattering (DLS) (Litesizer DLS 700, Anton Paar) using disposable polystyrene cuvettes at temperatures from 20 to 50 °C. For each temperature step, measurements were performed three times after equilibration for about 10 minutes.

Morphologies of Si@PNIPAM particles were examined using TEM (H-7000, Hitachi). TEM samples were prepared by drop-casting a Si@PNIPAM particle suspension on a TEM mesh and dried at ambient temperature overnight. Si@PNIPAM particles and Si NSs on a substrate were observed by SEM (JSM-IT800HL, JEOL) after coating with Osmium (Os) with the thickness of 9 nm. Dark-field scattering spectra of a single Si NS and Si@PNIPAM particle were measured by using a custom-built inverted optical microscope. In a reflection configuration, $SiO_2$ substrate was placed face-down onto the stage and illuminated by a halogen lamp *via* the same objective lens and transferred to the entrance slit of a monochromator (Kymera 328i, Andor) and detected by a cooled CCD (Newton, Andor).

Extinction spectra of diluted suspensions were measured using a double-beam spectrophotometer (UV-3101PC, Shimadzu). Specular reflectance spectra of solid samples were measured by a double-beam spectrophotometer (UV-3101PC, Shimadzu) with the incident and detection angles of 5°. The diffuse reflectance spectra with the incident angle of 0° was measured by a double-beam spectrophotometer equipped with an integrating sphere (SolidSpec 3700, Shimadzu). Reflectance of the area of 6 × 10 mm$^2$ within a 1 × 1.5 cm$^2$ size substrate was collected. In both measurements, the incident light was not polarized. Total reflectance spectra were obtained by adding the specular and diffuse reflectance spectra.

**Acknowledgements**

The authors thank Ms. Kana Kondo and Mr. Hiroki Kasai for their valuable assistance on the PNIPAM shell removal process. This work is partially supported by JST Commercialization Support, Grant Number JPMJSF2405, JSPS Bilateral Joint Research Project 120249928, and JSPS KAKENHI, Grant Numbers 24K01287 and 25K01608.

# Supporting Information

**Polymer-Shell Coating of Mie-Resonant Silicon Nanospheres for Controlled Fabrication of the Self-Assembled Monolayer**

Oanh Vu, Jialu Song, Hiroshi Sugimoto*, and Minoru Fujii*

Department of Electrical and Electronic Engineering, Graduate School of Engineering, Kobe University, Rokkodai, Nada, Kobe 657-8501, Japan

E-mail: sugimoto@eedept.kobe-u.ac.jp

E-mail: fujii@eedept.kobe-u.ac.jp

**S1. Extraction of average diameter of Si NSs and the distribution from the extinction spectrum**

We estimate the average diameter of Si NS core ($D_{\text{core}}$) and the standard deviation ($\sigma$) by fitting the scattering spectrum with calculated scattering cross-section spectrum assuming a normal size distribution. It is important to note that a PNIPAM shell does not affect the scattering spectrum of a Si NS in water because the refractive index is very close to that of water, i.e., scattering spectra of a bare Si NS and a Si@PNIPAM particle in water are very similar, and thus the diameter estimated from the extinction spectrum corresponds to that of the Si NS core in a Si@PNIPAM particle. **Figure S1** shows a measured extinction spectrum and the fitting result. The fitting reproduces the experimental spectrum very well.

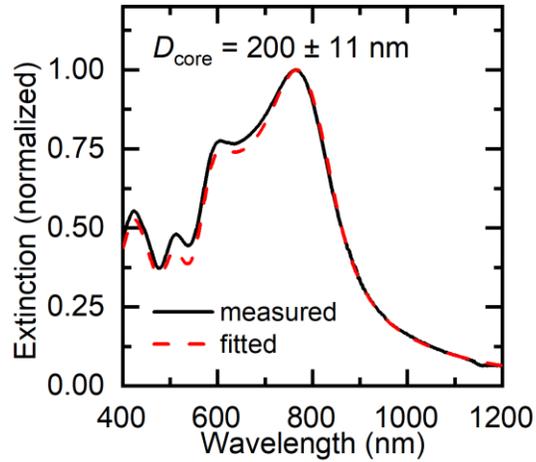

**Figure S1.** Normalized extinction spectrum of a Si@PNIPAM particle suspension (black) together with calculated scattering cross-section spectrum assuming a normal size distribution (red). The average diameter and the standard deviation obtained from the fitting is shown in the figure.

**S2. Extinction spectra of Si@PNIPAM particle suspension before and after a temperature cycle**

In **Figure 3d** in the main text, the extinction spectrum of a Si@PNIPAM suspension progressively changes by increasing the temperature from 20 °C to 50 °C. In **Figure S2**, the spectrum after heating up to 50 °C and then cooling down to 20 °C is shown together with that before the heating. The two spectra agree very well, indicating that this is the reversible process.

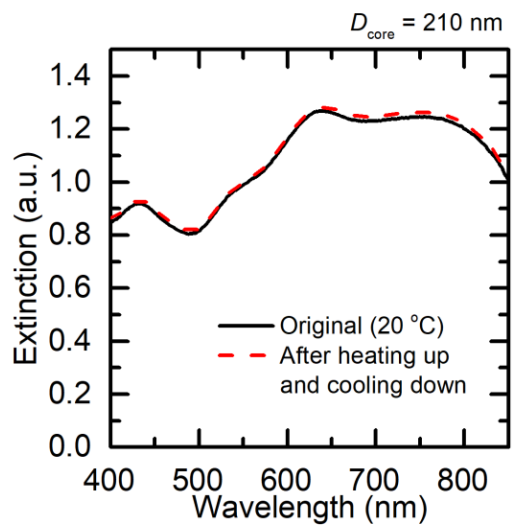

**Figure S2.** Extinction spectra of Si@PNIPAM particle suspension at 20 °C (solid black curve) and after heating up to 50 °C and cooling down to 20 °C (red dashed curve).

**S3. Temperature dependent property of Si@PNIPAM particle suspension with $D_{core}$=123 nm**

The data equivalent to **Figure 3** in the main text is shown for smaller particles ($D_{core}$=123 nm).

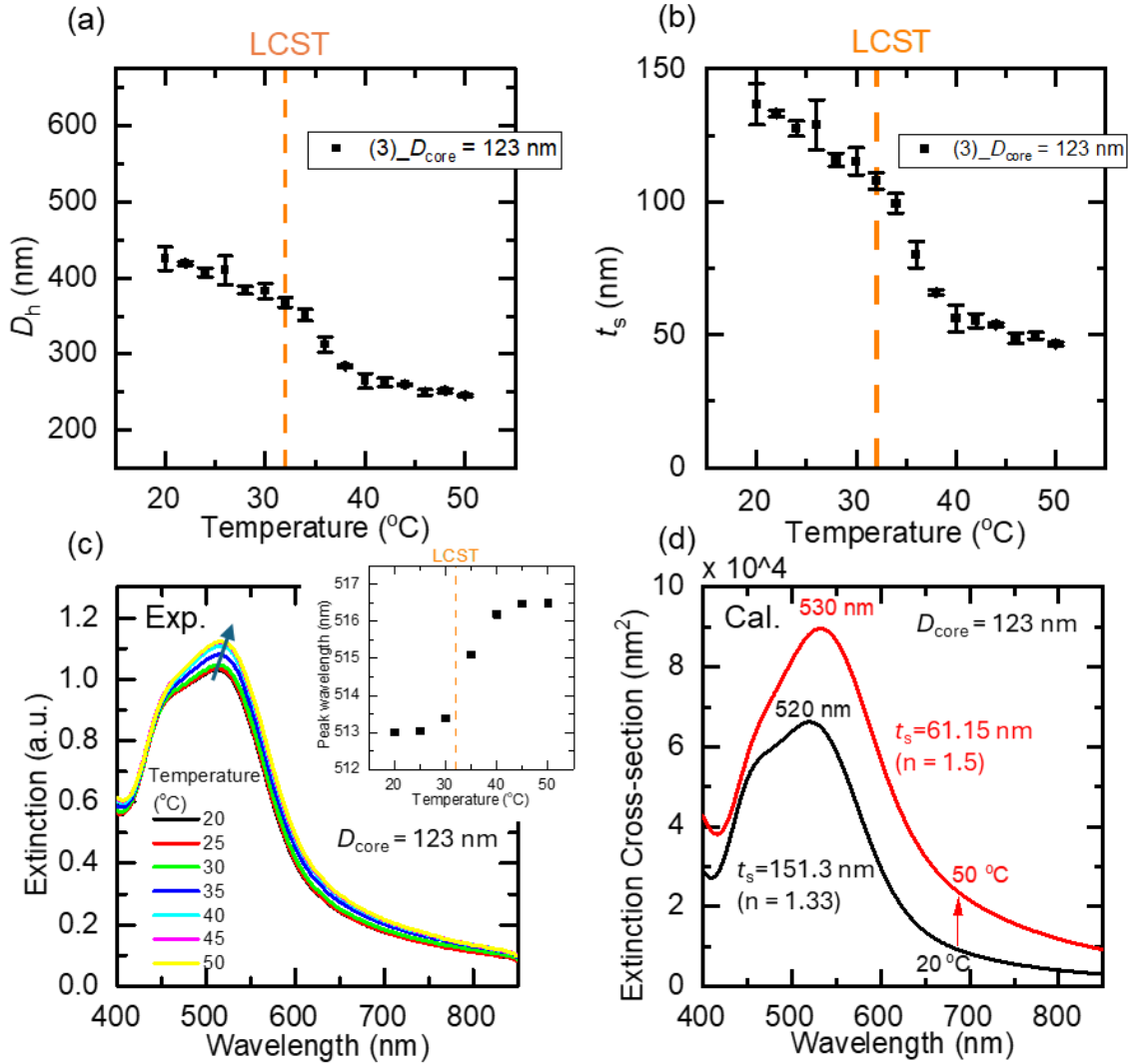

**Figure S3.** (a) $D_h$ of Si@PNIPAM particles ($D_{core}$=123 nm) fabricated with condition (3) as a function of the temperature. (b) $t_s$ as a function of temperature. (c) Temperature dependence of the extinction spectra. Inset: peak wavelength of the MD Mie resonance as a function of temperature. (d) Calculated extinction cross-section spectra of a Si@PNIPAM particle below and above the LCST. The water-rich phase below the LCST is modeled with the shell of the refractive index same as that of water ($n$ =1.33) (black curve), while the polymer-rich phase above LCST is modeled with a 151.3 nm-thick shell of $n$=1.5 (red curve). The shell thickness is estimated from experimental data.

## S4. TEM images of Si@PNIPAM particles having thin shells

When PNIPAM shells are very thin, it is not straightforward to distinguish whether Si NSs are physically attached or not from the SEM images. We therefore dropped Si@PNIPAM particle suspensions on a TEM mesh, dried and observed. The results are shown in **Figure S4**. PNIPAM shells are conformally formed on the surface of Si NSs, and thus Si NSs are not physically attached.

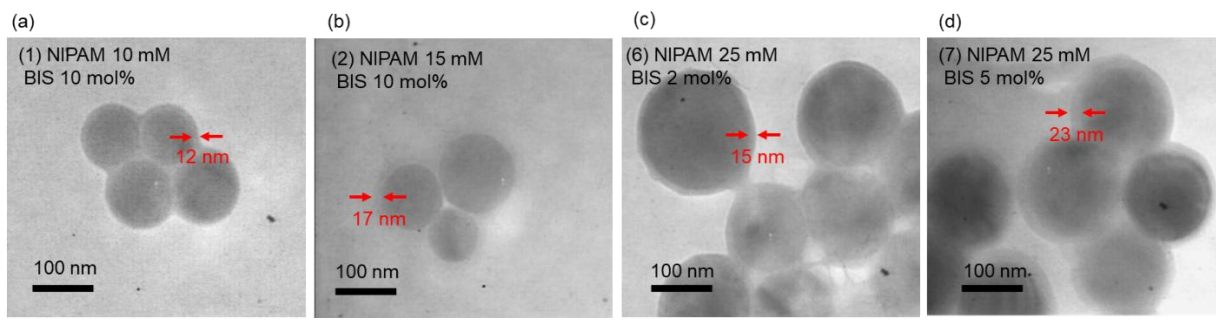

**Figure S4.** TEM images of dried Si@PNIPAM particles. The preparation conditions are (a) condition (1), (b) condition (2), (c) condition (6), and (d) condition (7).